\documentstyle[twoside,fleqn,espcrc2,psfig]{article}
\def\be{\begin{equation}}
\def\ee{\end{equation}}
\def\bea{\begin{eqnarray}}
\def\eea{\end{eqnarray}}

\def\aproxgt{\mathrel{%
      \rlap{\raise 0.511ex \hbox{$>$}}{\lower 0.511ex \hbox{$\sim$}}}}
\def\aproxlt{\mathrel{%
      \rlap{\raise 0.511ex \hbox{$<$}}{\lower 0.511ex \hbox{$\sim$}}}}

\newcommand{\AmS}{{\protect\the\textfont2
  A\kern-.1667em\lower.5ex\hbox{M}\kern-.125emS}}

\title{The Physical Interpretation of X-ray Phase Lags and Coherence: \\ RXTE
Observations of Cygnus X--1 as a Case Study}

\author{M.A. Nowak\address{JILA, Campus Box 440, Boulder, CO ~80309-0440, USA},
	J.B. Dove\address{Dept. of Physics and Astronomy, University of
	Wyoming, Laramie, WY ~82071, USA},    
	B.A. Vaughan\address{Space Radiation Lab, California
	Institute of Technology, 220-47 Downs, Pasadena, CA ~91125, USA},
	J. Wilms$^{\rm a,}$\address{IAA,
	Abt. Astronomie, Waldh\"auser Str. 64, D-72076 T\"ubingen, Germany}, 
	and M.C. Begelman$^{\rm a,}$\address{Dept. of APS, University of
	Colorado, Boulder, CO ~80309, USA}}
       
\begin{document}

\begin{abstract}
There have been a number of recent spectral models that have been
successful in reproducing the observed X-ray spectra of galactic black hole
candidates (GBHC).  However, there still exists controversy over such
issues as: what are the sources of hard radiation, what is the system's
geometry, is the accretion efficient or inefficient, etc.  A potentially
powerful tool for distinguishing among these possibilities, made possible
by the {\it Rossi X-ray Timing Explorer (RXTE),} is the variability data,
especially the observed phase lags and variability coherence.  These data,
in conjunction with spectral modeling, have the potential of determining
physical sizes of the system, as well as placing strong constraints on both
Compton corona and advection models.  As an example, we present {\it RXTE}
variability data of Cygnus X-1.
\end{abstract}

\maketitle

\section{INTRODUCTION}

Many black hole candidate (BHC) systems show three or more spectral states:
the ``low'' (X-ray hard), ``high'' (X-ray soft), and ``very high'' (X-ray
soft, plus a hard, power-law tail) states.  Typically, sources are seen to
be in the low state when they are below $\sim 10\%$ of their Eddington
luminosity, and they tend to exist in the high and very high states at
observed luminosities above $10\%~L_{\rm Edd}$ \cite{nowak}.  In this
article, we will concentrate on the properties of the low state, for which
we have obtained $\sim 20$ ks of Cyg X-1 data with the {\it Rossi X-ray
Timing Explorer (RXTE)} \cite{dovec,doved}.

A variety of models have been proposed for the low state of BHC in general,
and for the low state of Cyg X-1 in specific.  Currently, these models fall
into two main classes: the Compton corona models
\cite{dovec,doved,tit,haardt,gier,dovea,doveb,pout} and the Advection
Dominated Accretion Flow (ADAF) models \cite{nar,esin}.  For the Compton
corona models, the currently favored geometry is one in which a central,
spherical corona is surrounded by an exterior, cold disk
\cite{dovec,doved,gier,dovea,doveb,pout}. This is similar to the geometry
proposed by the ADAF models \cite{nar,esin}; however, the ADAF models
invoke different sources for the Comptonization seed photons\footnote{ADAFs
take the Comptonization seed photons to be cyclo/synchrotron photons
generated within the advective region itself.}, and they make more detailed
predictions for the dynamics of the inner region.  Specifically, the
advective inner region is moving quasi-radially, at close to the free-fall
velocity, toward the black hole.

We note that Cyg X-1 has upon occasion, including very recently, transited
to the high state \cite{cui}.  Different scenarios have been proposed
for this transition in both the Compton corona picture \cite{pout} and the
ADAF picture \cite{esin}.  We will not discuss this transition here.
Instead, we will concern ourselves with how variability data, i.e.
the phase/time lags and the coherence function, can be used to constrain
models within a given state.  As described below, these data currently pose
challenges for both the Compton corona and ADAF models of the low state.

\section{SPECTRAL MODEL OF CYG X-1}\label{sec:compt}

As an example of a successful (in terms of fitting the energy spectra) low
state model, we consider the spherical corona model of Dove et
al. \cite{dovea,doveb,dovec,doved}.  Its basic features are as follows. The
inner region of the accretion system is modeled as a spherical cloud with
total optical depth $\sim 1.6$ and an {\it average} temperature of $87$
keV.  This spherical region is then surrounded by a cold ($T \le 150$ eV),
geometrically thin disk which reprocesses hard radiation from the corona,
as well as provides seed photons for Comptonization.  The overall quality of
the fit of this model to our Cyg X-1 data is quite good ($\chi^2_{\rm red}
= 1.6$).  We will use this as our ``straw man'' model when we discuss 
the limits that phase lags place on coronal models (\S\ref{sec:compphase}).

\section{POWER SPECTRAL DENSITY}

The power spectral density (PSD) is calculated by taking lightcurves from
the data, dividing them into segments of equal length, and then taking the
Fast Fourier Transform (FFT) of each data segment.  The squared amplitude
of each individual FFT (for a given lightcurve) is then averaged together.
Furthermore, we usually average over (logarithmically spaced) Fourier
frequency bins as well.  This yields the resulting PSD for each lightcurve.
Here we choose a normalization such that the integral of the PSD over
Fourier frequency yields the {\it square} of the total {\it root mean
square (rms) amplitude} of the variability \cite{miya}.  For a given narrow
Fourier frequency interval, the rms is (to within a factor of $\sqrt{2}$) the
fractional amount by which the lightcurve is sinusoidally modulated in that
given frequency interval.

\begin{figure}
\centerline{
\psfig{figure=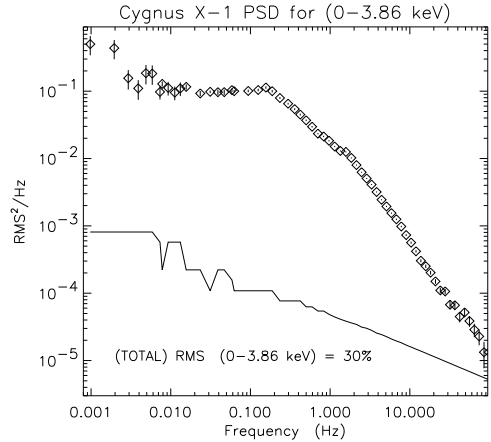,width=0.48\textwidth}
}
\vskip-0.2 true in
\caption{Power Spectral Density of the (0-3.86 keV) band of Cyg X-1.  The
solid line corresponds to the ``effective'' noise level. For the
normalization of Miyamoto \cite{miya}, this is 2/(Count Rate), divided by
the number of FFTs and frequency bins averaged over.  FFTs for lightcurves
of 1024, 128, and 32 s duration were combined for this plot.}
\label{fig:psd}
\end{figure}

In Figure \ref{fig:psd} we plot the PSD of the low energy band ($0-3.86$
keV) for our Cyg X-1 data.  This PSD, in both shape and amplitude, is very
similar to what has been previously observed.  We note, however, that there
is a weak, broad feature at $0.005$ Hz, with rms $\sim 1.5\%$.  A
Lomb-Scargle periodogram \cite{scarg} indicates that the significance of
this feature is $\sim 50\%$.  However, for reasons discussed in
\S\ref{sec:coher}, this feature may be somewhat more significant than that.

\section{PHASE LAGS}

\subsection{Basics}\label{sec:basics}

\begin{figure}
\centerline{
\psfig{figure=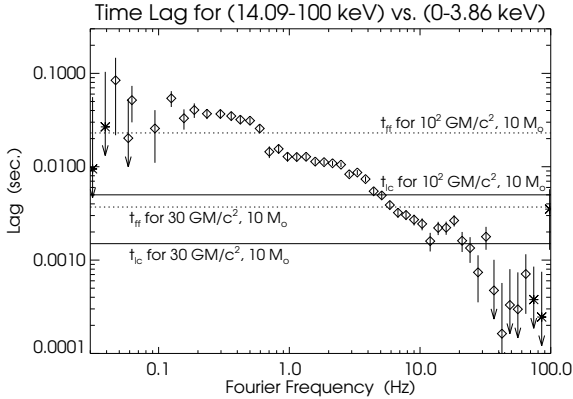,width=0.48\textwidth}
}
\vskip-0.2 true in
\caption{Time lags (as a function of Fourier frequency) in Cyg X-1 for the
($14.09-100$ keV) vs. ($0-3.86$ keV) energy bands.  Open diamonds, the hard
lags behind the soft; $*$ the soft lags behind the hard.  Horizontal lines
are characteristic timescales for a 10 $M_\odot$ black hole.  Dotted lines
are the (radial) free-fall timescales from $R=30, ~100~GM/c^2$, and solid
lines are the light crossing times for $R=30, ~100~GM/c^2$.}
\label{fig:times}
\end{figure}

Fourier phase lags (and equivalently Fourier time lags) are calculated from
the FFTs constructed from data segments of two different, but concurrent,
lightcurves.   Let $s(t)$ be a ``soft energy'' light curve and $h(t)$ be a
``hard energy'' lightcurve.  The Fourier phase lag, $\phi(f)$, is then just the
phase of the (complex) quantity
\begin{equation}
\langle S^*(f) H(f) \rangle ~~,
\label{eq:lag}
\end{equation}
where $S(f)$ and $H(f)$ are the FFTs of the concurrent data segments of
$s(t)$ and $h(t)$, respectively.  The $^*$ denotes a complex conjugate and
the angle brackets denote an average over data segments and/or
Fourier frequency bins.  The time lag, $\tau(f) \equiv
\phi(f)/2\pi f$.

In general, both the phase lag and time lag are non-constant functions of
Fourier frequency, $f$.  However, we naively expect that the {\it longest}
observed time lag in the system is no longer than the {\it longest}
distance in the system divided by the {\it slowest} propagation speed.  As
shown in Fig. \ref{fig:times}, this longest time for Cyg X-1 is $\sim 0.1$
s.  This is considerably longer than both the expected free-fall timescales for
ADAF models and the sound travel time of hot corona models.  Therefore,
these data may pose a problem for both of these models.  

Similarly, the {\it shortest} observed time lag should be longer than the
{\it smallest} scale (over which significant luminosity is generated)
divided by the {\it fastest} propagation speed. As shown in
Fig. \ref{fig:times}, the shortest time lags observed in Cyg X-1 are
comparable to a light crossing time, and therefore may be relevant to
determining physical parameters for Compton corona models
(cf. \S\ref{sec:compphase}).

If the coherence function is unity (cf. \S\ref{sec:coher}), one can often
think of the phase lag in terms of a transfer function.  For this case, we
have $H(f) = A(f) \exp[i \phi(f)] S(f)$, where $A(f)$ is a real-valued
amplitude.  In the time domain, this means that $h(t)$ is the convolution
of $s(t)$ with a {\it linear} transfer function.  Specifically, 
$$h(t) = \int_{-\infty}^{\infty} t_r(t - \tau) s(\tau) ~d\tau ~~.$$
The Fourier transform of $t_r(\tau)$ is then just $A(f) \exp[i \phi(f)]$.

\subsection{Simple Interpretations}\label{sec:simpint}

Imagine that we have a source of fluctuations that produces soft
X-ray photons that is some distance from a region that produces hard X-ray
photons.  If the disturbances can propagate from the soft photon producing
region to the hard photon producing region--- without dispersion--- then we
expect there to be a time delay between the soft and hard photons that is
{\it independent of Fourier frequency}, $f$.  The time delay at all Fourier
frequencies will simply be the distance between the soft X-ray source and
the hard X-ray response divided by the propagation speed. This means that
the Fourier phase lag will increase linearly with $f$ (modulo integer
multiples of $2\pi$).

As shown in Fig. \ref{fig:times}, the time lags observed in Cyg X-1 are not
independent of Fourier frequency.  If anything, the Fourier phase lag is
more approximately independent of $f$.  A crude, but illustrative, model of
this is as follows.  If $H(f)$ is the transform of the hard photon light
curve and $S(f)$ is the transform of the soft photon light curve, let
\begin{equation}
H(f) ~=~ A \exp(\pm i \Delta \phi) ~ S(f) ~ ~,
\end{equation}
where both $A$ and $\Delta \phi$ are constants, and the $+$ is for $f > 0$
and the $-$ is for $f < 0$.  [The antisymmetry in the relative phase is due
to the fact that real light curves produce Fourier transforms where $H(f) =
H^*(-f)$.] Taking the inverse transform of $A \exp(\pm i \Delta \phi)$,
the transfer function, $t_r(\tau)$, can be written in the time domain as
\begin{equation}
t_r(\tau) ~=~ A \left [ \cos(\Delta \phi)
\delta(\tau) + {{\sin(\Delta \phi)}\over{\tau}} \right ] ~~.
\end{equation}
That is, a fraction $A \cos(\Delta \phi)$ of the hard variability is
{\it exactly} coincident with the soft variability, while a (typically smaller)
fraction is delayed from the soft variability with a $\tau^{-1}$ ``tail''
in the transfer function.  For $A\sim 1$ and $\Delta \phi \sim 0.1$
radians, roughly 90\% of the soft and hard lightcurves are exactly
coincident with one another.

\subsection{Simple Propagation Models}\label{sec:prop}

It is often convenient to think of propagating disturbances in terms of
sources and responses.  For example, we might have a disturbance,
$\Psi(\vec x, t)$ that obeys a wave equation
\be
\left ( {{\partial^2}\over {\partial \vec x~  ^2}} ~-~
c_p^{-2} {{\partial^2}\over {\partial t ^2}}  \right )  \Psi(\vec x, t)
~~=~~  -4 \pi \rho_D(\vec x, t) ~,
\label{eq:wave}
\ee
where $\rho_D(\vec x, t)$ is the source of the disturbance. 

Imagine that the source of fluctuations, $\rho_D$, is separable in
space and time.  Let us also assume that the {\it observed} soft X-ray
light curve is equal to the disturbance, $\Psi$, times a weighting
function, $g_s(\vec x)$, integrated over the system.  Likewise, let us
assume that the {\it observed} hard X-ray light curve is related to $\Psi$
via a weighting function $g_h (\vec x)$.

As was discussed by Vaughan \& Nowak \cite{vam}, it is relatively
straightforward to calculate the resultant phase lags for this case.
Furthermore, the resultant ``transfer function'' between the soft and hard
lightcurves has many of the qualitative properties discussed in
\S\ref{sec:simpint}.

\begin{figure}
\centerline{
\psfig{figure=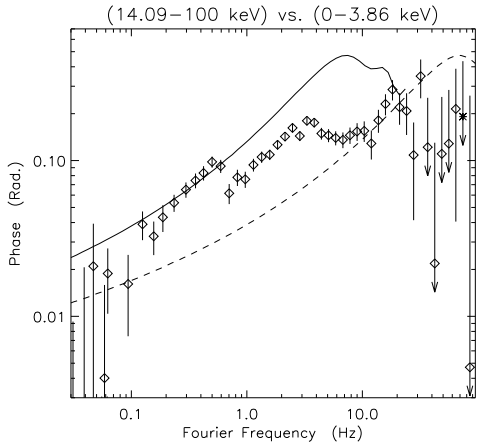,width=0.48\textwidth}
}
\vskip-0.2 true in
\caption{Phase lags (radians) for the data presented in
Fig. \ref{fig:times}. Lines correspond to the simple propagation model of
\S\ref{sec:prop}.  The solid line is for $c_p \approx 0.01 ~c$, and the dashed
line is for $c_p \approx 0.1 ~c$, where $c$ is the speed of light.}
\label{fig:green}
\end{figure}

As a simple example, let us consider the (phenomenological) weighting functions:
\bea
g_s(r) &\propto& \theta \left ( {{r} \over {r_0} } - 6 \right ) *
               \exp \left ( -\alpha_s {{r}\over{r_0}} \right ) ~~,~~
        \nonumber \\
g_h(r) &\propto& \theta \left ( {{r} \over {r_0} } - 6 \right ) *
               \exp \left ( -\alpha_h {{r}\over{r_0}} \right ) ~~,
\eea
where $-\alpha_s$ and $-\alpha_h$ are constants, and $\theta$ represents a
step function.  Here we shall take
$\alpha_s = 0.31$ and $\alpha_h = 0.58$, so that the response of the soft
X-rays is predominantly at $r \aproxlt 15 ~r_0$ and the response of the
hard X-rays is predominantly at $r \aproxlt 8 ~r_0$.  
For the case of Cyg X-1, we will take $r_0
= 6 ~GM/c^2$ with $M= 10~M_\odot$.  The resultant phase lags then depend
upon $c_p$, the propagation speed of the disturbances (which we take to be
uniform, with the waves directed toward a ``sink'' at the inner disk edge)
\cite{vam}. 

As shown in Fig. \ref{fig:green}, such a model qualitatively reproduces the
observed phase lags.  However, to obtain quantitative agreement with the
phase lags at low Fourier frequency ($\sim 0.1$ Hz), a very slow
propagation speed of $c_p \approx 0.01 ~c$, where $c$ is the speed of
light, is required.  As discussed in \S\ref{sec:basics}, such a slow speed
is much less than both the radial velocity in ADAF models, as
well as the sound speed for Compton corona models, and therefore is
problematic for both models.

\subsection{Comptonization Models}\label{sec:compphase}

Any {\it intrinsic} time delays between soft and hard photons for
Comptonization seed photons will be modified as the seed photons diffuse
through the Compton cloud \cite{mi88,mill,mav}. The greater the observed output
energy, the longer the diffusion time through the Compton corona.  As has
been discussed by Miller \cite{mill} and Nowak \& Vaughan \cite{mav}, the
{\it minimum} expected time delay is the difference in the diffusion time
for the {\it observed} hard and {\it observed} soft photons.

\begin{figure}
\centerline{
\psfig{figure=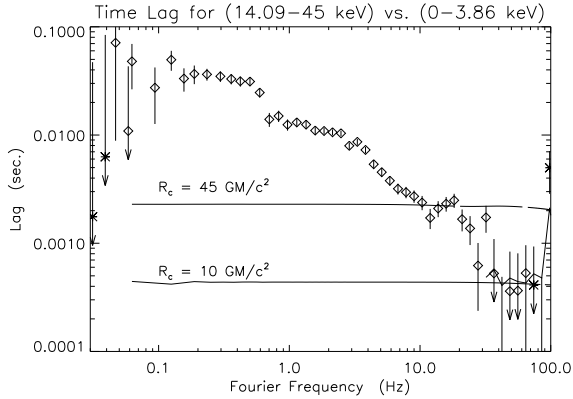,width=0.48\textwidth}
}
\vskip-0.2 true in
\caption{Time lags (as a function of Fourier frequency) for Cyg X-1 data.
The solid lines correspond to the {\it minimum} expected time lag for a
spherical Compton corona (with parameters given in \S\ref{sec:compt}) of
radius $R = 45 ~GM/c^2$ and $R = 10 ~GM/c^2$ $(M=10~ M_\odot)$. Thick 
line is the magnitude of the Poisson noise limit above 30 Hz.}
\label{fig:shelf}
\end{figure}

We have calculated this minimum time delay for the Compton corona model
discussed in \S\ref{sec:compt}, and show this along with the data in
Fig. \ref{fig:shelf}.  Depending upon whether one identifies the $\sim
10-30$ Hz time lags or the $\aproxgt 30$ Hz time lags as the upper limit
for such a time lag ``shelf'', an {\it upper} limit to the coronal radius
is seen to be $R \approx 45 ~GM/c^2$ or $R \approx 10 ~GM/c^2$
($M=10~M_\odot$), respectively.  For either case, such a small radial
extent of the corona poses a severe problem for ADAF models, as shown in
Fig. \ref{fig:times}, which would then have great difficulty explaining the
longest time lags.  (There still remains the possibility for Compton
corona models that the longest time lags are ``intrinsic'' time lags from
the outer cold disk, merely ``reprocessed'' by the corona \cite{mill,mav}.)

To be conservative, we identify the $10-30$ Hz data as the upper limit for
the shelf, as the $>30$ Hz data shows a strong loss of coherence (cf. Fig
\ref{fig:coher}). We clarify this point below where we describe exactly
what we mean by the coherence function.

\section{COHERENCE FUNCTION}\label{sec:coher}

The coherence function has been extensively discussed by Vaughan \& Nowak
\cite{vam}.  Essentially, it is the average, normalized amplitude of the
cross power spectral density (which is what one is computing to find the
time lags).  Similar to eq.~\ref{eq:lag}, the coherence function, $C(f)$ is
given by
\be
C(f) \equiv {{\langle S^*(f) H(f) \rangle^2} \over { \langle S^2(f) \rangle
\langle H^2(f) \rangle }} ~~.
\ee
Various methods exist for determining, and minimizing, the effects of
Poisson noise on the estimate of this function \cite{vam}.

\begin{figure}
\centerline{
\psfig{figure=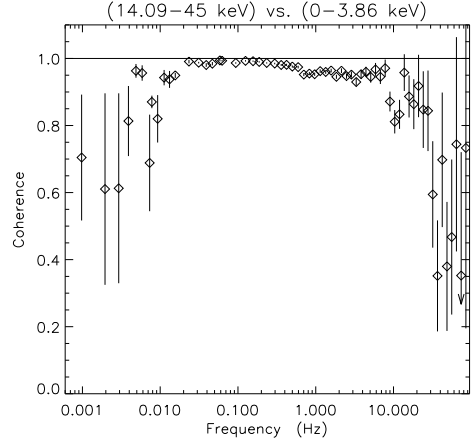,width=0.48\textwidth}
}
\vskip-0.2 true in
\caption{Coherence function (as a function of Fourier frequency) for the
data presented in Fig. \ref{fig:times}.  Note how remarkably close the
coherence is to unity in the range $\sim 0.02 - 10$ Hz. The loss of
coherence at frequencies below $\sim 0.02$ Hz is real, as is the recovery
to near unity coincident with the $0.005$ Hz feature from
Fig. \ref{fig:psd}.  The loss of coherence at high frequency ($\aproxgt 30$
Hz) may be influenced by uncertainties associated with the Poisson noise
level and the instrumental deadtime.}
\label{fig:coher}
\end{figure}

As also discussed by Vaughan \& Nowak \cite{vam}, most previously
considered mechanisms for producing variability in BHC systems
(i.e. multiple shots or flares, any ``nonlinear'' processes, etc.) will
lead to a strong {\it loss} of coherence.  This makes the near unity
coherence shown in Fig. \ref{fig:coher} quite remarkable.  The near unity
coherence seen between $\sim 0.02 - 10$ Hz argues for a single source of
disturbances and/or a global response in Cyg X-1 over (these rather
disparate) timescales \cite{vam}.

In a very real sense, the coherence function is a measurement of the degree
of constancy of the phase lags between soft and hard photons as one
averages over individual data segments and Fourier frequency bins (as
discussed in \S\ref{sec:basics}).  If the phase is constant from data
segment to data segment, the coherence is unity. A variable phase, as might
be caused by two or more uncorrelated processes relating soft and hard
photons, will lead to coherence loss \cite{vam}.  As the coherence drops
rapidly above $\aproxgt 30$ Hz, we cannot be sure that the observed time
lag in this regime is indicative of {\it solely} the diffusion through a
Compton corona.  For this reason, we take the $10-30$ Hz data as the
conservative estimate.

Note also that there is a definite loss of coherence below $\sim 0.02$ Hz.
There is a clear trend which is only broken near $0.005$ Hz.  At that
point, there is a recovery of coherence to near unity, coincident with the
feature seen in the PSD (cf. Fig. \ref{fig:psd}).  This indicates that,
although the Lomb-Scargle periodogram indicated a significance of only
50\%, this feature may actually be significant.  At the very least, it
argues that PSDs of Cyg X-1 should be extended to frequencies as low as
$~10^{-3}$ Hz whenever statistics permit.

\section{SUMMARY}

We have discussed the role of Fourier phase/time lags and variability
coherence in constraining spectral models of BHC.  Much of the theory
behind this work can be found in the papers of Dove et
al. \cite{dovea,doveb}, Miller \cite{mill}, Nowak \& Vaughan
\cite{mav}, and Vaughan \& Nowak \cite{vam}.  To illustrate our points, we
have presented {\it RXTE} data of Cyg X-1 (cf. \cite{dovec,doved}).

Despite the general success of spectral models, such as the Compton corona
model discussed in \S\ref{sec:compt}, the variability data pose
substantial challenges.  For example, both the ADAF models and Compton
corona models (if they rely on sound speed propagation timescales) have
substantial difficulty in explaining the longest observed time lags.
Explaining the observed unity coherence in Cyg X-1 will be difficult for
all models unless they can postulate a {\it global} variability mechanism.

The shortest observed time lags were also seen to provide upper limits to
the size of a spherical Compton corona.  The ``recovery'' to near unity
coherence at $\sim 0.005$ Hz, coincident with a weak feature in the PSD,
indicates the possibility of an extremely low frequency oscillation in Cyg
X-1.

In short, no currently proposed model for the low state of BHC in general,
or the low state of Cyg X-1 in specific, can claim to completely describe
both the spectral and variability data.  However, with the advent of
{\it RXTE} and {\it BeppoSAX}, which between them are
capable of broad spectral coverage and fast timing, only models that can
explain, or at least be consistent with, both the spectral and variability
data should be considered to be viable.


\begin{thebibliography}{9}
\bibitem{nowak}	M.A. Nowak, PASP 718 (1995) 1207.
\bibitem{dovec} J.B. Dove, J. Wilms, M.A. Nowak, B. Vaughan, M.C. Begelman,
	        MNRAS (in Press) (1998) (astro-ph/9707322).
\bibitem{doved} J.B. Dove, M.A. Nowak, J. Wilms, B. Vaughan, this Volume 
		(1998).
\bibitem{tit}	L. Titarchuk, ApJ 434 (1994) 570.
\bibitem{haardt} F. Haardt, L. Maraschi, \& G. Ghisellini, ApJ 476 (1996)
		670.
\bibitem{gier}  M. Gierli\'nski, et al., MNRAS 288 (1996) 958.
\bibitem{dovea} J.B. Dove, J. Wilms, M.C. Begelman, ApJ 487 (1997) 747.
\bibitem{doveb} J.B. Dove, J. Wilms, M.G. Maisack, M.C. Begelman, ApJ 487 
	        (1997) 759. 
\bibitem{pout}  J. Poutanen, J.H. Krolik, \& F. Ryde, MNRAS (Submitted)
		(1997) (astro-ph/9709113).
\bibitem{nar}	R. Narayan \& I. Yi, ApJ 428 (1994) L13.
\bibitem{esin}	A. Esin, R. Narayan, W. Cui, J.E. Grove, \& S.N. Zhang,
		ApJ, Submitted (astro-ph/971167). 
\bibitem{cui}	W. Cui, S.N. Zhang, W. Focke, \& J.H. Swank, ApJ 484 (1997)
		383.	
\bibitem{miya} 	S. Miyamoto, S. Kitamoto, S. Iga, H. Negoro, \& K. Terada,
		ApJ 391 (1992)  L21.
\bibitem{scarg} J.D. Scargle, ApJ 263 (1982) 835.
\bibitem{vam} 	B.A. Vaughan \& M.A. Nowak, ApJ 474 (1997) L43.
\bibitem{mi88}	S. Miyamoto, S. Kitamoto, K. Mitsuda, \& T. Dotani, Nature
		336 (1988) 450.
\bibitem{mill}	M.C. Miller, ApJ 441 (1995) 770.
\bibitem{mav}	M.A. Nowak \& B.A. Vaughan, MNRAS 280 (1996) 227.


\bigskip
We would like to acknowledge useful conversations with P. Michelson, M. van
der Klis, D. Gruber, and K. Jahoda.  This work was supported in part by
NASA grants NAG 5-3225 and NAG 5-3310 (MAN), as well as NASA Grants NAG
5-2026, NAG 5-3239, NSF Grants AST91-20599, AST95-29175, INT95-13899, DARA
grant 50 OR 92054, and by a travel grant to J.W. from the DAAD.


\end{thebibliography}
\end{document}